# STRATEGIES IN EDUCATION, OUTREACH, AND INCLUSION TO ENHANCE THE US WORKFORCE IN ACCELERATOR SCIENCE AND ENGINEERING*

M. Bai (SLAC), W.A. Barletta (MIT), D.L. Bruhwiler (RadiaSoft LLC), S. Chattopadhyay (FNAL/NIU), Y. Hao (MSU/BNL), S. Holder (SLAC), J. Holzbauer (FNAL), Z. Huang (SLAC), K. Harkay (ANL), Y.-K. Kim (UChicago & CBB), X. Lu (NIU/ANL), S.M. Lund (MSU/USPAS), N. Neveu (SLAC), P. Ostroumov, (MSU), J. R. Patterson (Cornell/CBB), P. Piot (NIU/ANL/CBB), T. Satogata (JLab), A. Seryi (JLAB/ODU), A.K. Soha (FNAL), S. Winchester (USPAS/FNAL)

*Abstract*

We summarize the community-based consensus for improvements concerning education, public outreach, and inclusion in Accelerator Science and Engineering that will enhance the workforce in the USA. The improvements identified reflect the product of discussions held within the 2021-2022 Snowmass community planning process by topical group AF1: *Beam Physics and Accelerator Education* within the *Accelerator Frontier.* Although the Snowmass process centers on high-energy physics, this document outlines required improvements for the entire U.S. accelerator science and engineering enterprise because education of those entering and in the field, outreach to the public, and inclusion are inextricably linked.

## 1. EXECUTIVE SUMMARY

**To be authored by Lund, Bai, and Huang after composition of the main document. Due to timing issues this may be done as an update after the initial document check-in on March 15, 2020.**

**Accelerator Frontier leaders have advised that Snowmass white papers should have an executive summary to maximize impact in the final Snowmass report and the following P5 process.**

## 2. INTRODUCTION

> In 1812, presenting his Elements of Chemical Philosophy, Sir Humphrey Davy wrote: *"Nothing tends so much to the advancement of knowledge as the application of a new instrument"*. Galileo Galilei (1564 – 1642) himself, professed to us much earlier already: *"Measure what is measurable and make measurable what is not so"*.

No instrument could be more appropriate for these pronouncements than charged particle accelerators. Accelerators are notable in the sense that they are a key tool of discovery science and high technology while the science and technology of accelerators is itself a developed scientific discipline. Accelerators are crucial to the future health of US programs in the discovery sciences which heavily rely on the large accelerator facilities in high energy physics, nuclear physics, and photon and neutron sciences that are vital tools in materials science, chemistry, and biology. Accelerator systems, both large and small, also play a crucial role in a wide and ever-growing range of applications in high-technology commerce, medicine, clean-energy, and national security. Technological components and guiding concepts in accelerator systems for these disparate and important applications share a high degree of commonality. It is vital for the future health of both US discovery sciences and high technology that we recruit, train, and inspire a strong technical workforce to support the field.

Because of the interdisciplinary nature of Accelerator Science and Engineering (AS&E) and the relatively low representation of the field in university faculty, undergraduates from research universities are unlikely to be drawn into AS&E for specialized postgraduate training, and even less likely to have acquired core components of the background, knowledge base, and expertise demanded by the field. Hence those of us in the community must train our own to secure the future of our field. In the case of accelerators, we have an added historical precedent and responsibility of transferring the emerging technologies to the private sector for the benefit of society at large. Scientists working with accelerators will also benefit from more intimate knowledge of the working of their instruments.

Historically, high-energy and nuclear physics first motivated the advances in cyclotrons, synchrotrons and linear accelerators. Today the science and technology of accelerators is driven broadly with advances in other applications as likely to help future accelerator collider concepts for high energy physics as vice-versa. Within the US Department of Energy (DOE), the Office of Science has charged its Office of High Energy Physics (OHEP) to be the steward of the broad field of accelerator and technology. In this context, we summarize our Snowmass recommendations for the full field.

Improving AS&E components in: education to train strong future performers, outreach and recruiting to secure talent, and addressing diversity and inclusion issues for a healthy balance, all represent sound investments to better secure our technological future. We recommend that the DOE OHEP, together with NSF, should continue and extend broad-based support of AS&E issues in education, outreach, and inclusion to enhance the workforce. In this white paper, community recommendations for enhancements are summarized for education (**Sec. 3**), outreach (**Sec. 4**) and inclusion (**Sec. 5**). These sections are designed to be relatively self-contained.

## 3. EDUCATION

Resources required to improve AS&E education are relatively modest and cost effective. The US educational effort presently centers around university and (sometimes university affiliated) national laboratory programs drawing in graduate students and early career scientists and engineers to the field. The students and junior scientists are mentored by established professionals who nurture their development and training within projects and/or by directed study. Augmented training for graduate students and early career professionals is provided by the US Particle Accelerator School (USPAS) via two intensive school sessions per year. The curriculum of the USPAS covers the full field and evolves as the field changes under the guidance of community-derived governance committees. Recently, student support has been augmented by four DOE-funded traineeships at AS&E universities. Each traineeship is designed to start 4-6 graduate students per year in key areas of AS&E until they transition to national laboratories and other programs to complete MS or PhD thesis research.

In the sections below, we outline issues and opportunities for advancement in General Education (**Sec. 3.1**), Graduate Education in Universities (**Sec. 3.2**), the DOE Traineeships (**Sec. 3.3**), and the US Particle Accelerator School (**Sec. 3.4**).

### 3.1 General Education

The entire enterprise of accelerator-based discovery science, technology, medicine and commerce is enabled by a highly trained workforce, experts in the physics and engineering of particle beams, their associated instruments, methods, and technologies all implemented through the efforts of specialized technicians, machine operators, hardware designers, and production experts. Taken in total, the knowledge base underpinning the largest accelerator infrastructures, e.g., high-power and precision electromagnetics, industrial scale cryogenic and vacuum systems, precise positioning and alignment systems, complex integrated software controls and specialized electronics transcend the offerings of any single university's graduate curriculum. Yet, both initial and specialized continuing education are crucial to maintain the long-term health of the US accelerator workforce. Consequently, universities plus the public and private sector beneficiaries of accelerator science and technology should cooperate in building the accelerator workforce of the future. Also, for the universities to function effectively in their roles to recruit and launch future talent to support the field, they must have projects, grants, and facilities to support faculty and draw in students. Issues relating to this issue and community-identified opportunities for improvements are detailed in **Sec. 3.2**.

The resources required to maintain an adequate, well-trained workforce to deliver the technology for the discovery science, medicine, commerce, and national security sectors of the American economy are only a tiny fraction of the overall, multi-billion-dollar annual expenditures in these areas. Educational investments of the order of $5M annually can easily be structured to be cost effective[S3-1,2]. Optimally, these investments should be spread over three distinct segments of the accelerator workforce: trained technicians, BS and MS-level operations and design engineers and accelerator operators, and PhD-level accelerator physicists and research engineers. All segments also require outreach for maximum effectiveness (see **Sec. 4**) and efforts should be structured to improve persistent issues within AS&E with diversity, equity, and inclusion (see **Sec. 5**). The community would also benefit from public outreach efforts aimed at the nation's secondary schools, to increase awareness of the benefits conveyed by the field to society and inspire the generations to come to participate.

*3.1.1 Technicians:* Technicians play a vital role in accelerator projects and facilities and have become increasingly difficult to hire. In spite of this, negligible attention has been paid by the DOE and NSF to attract and train the accelerator technician workforce in community colleges and vocational training programs. Improved training and mission awareness within this sector of the workforce should have the attention of the Accelerator R&D and Production program office in partnership with the Office of Workforce Development for Teachers and Scientists. Both national laboratories and companies that supply accelerator components would benefit. General training strategies and rubrics could best be developed by convening a focused workshop under the joint sponsorship of the DOE/SC Accelerator Stewardship and Accelerator Development programs [S3-3] that would bring together engineering and accelerator operations managers with a representative group of educators from community colleges and vocational training centers.

*3.1.2 BS & MS engineers, designers and accelerator operators:* This group is a mainstay of hardware design, development and procurement, project engineering, and efficient operations at DOE national laboratories as well as in the commercial sector. A valuable outreach activity would be a DOE-supported program of seminars and colloquia given by lead engineers from national laboratories and presented in engineering departments of leading engineering universities in the US. The program should be aimed at awakening the interests of engineering students in the many challenges offered by accelerator-related research, production and operations programs in the national laboratories and in the commercial sector. Ideally lectures would be accompanied by recruiting sessions at the universities. These could naturally integrate with recruiting for MS programs in the DOE traineeships (**Sec. 3.3**).

Another missing component to reach this group as well as recruit is to revise and improve AS&E tutorials making the level lower for a gentle orientation to the field. Well-done tutorials can also be used as topical introductions and to recruit students into more comprehensive, in-person intensive courses (see **Sec. 3.4**) in the USPAS. One approach to filling this need and broadening scope of potential uses is for the US community to produce and maintain a set of targeted MOOCs (Massive Open Online Courses)[S3-4]. These could be free, on-demand and start anytime with embedded quizzes, and should be at the undergraduate level. MOOCs are not inexpensive (~$200K to $250K per course) to produce when made to production values at Coursera and U2 standards (MOOC platforms). However, quality MOOCs could fill in a persistent need for augmented training and refresher courses for engineers and BS-level physicists already in the accelerator workforce. A strategy to implement this would be best developed by a community workshop where a framework for approach, maintenance, and updates could be established, and the most relevant focus to start could be identified. Recommended initial topics include: 1) *Introduction to Accelerator Physics*, 2) *Basic Accelerator Technologies*, and 3) *Applications of Accelerators*. If the promise of this vision is proven, topics and scope could be generalized to reach additional parts of the workforce such as the managers in national laboratories with topics like *Project Management.* A developed example of an AS&E MOOC in the Coursera framework can be found in the Nordic Particle Accelerator Program's *Introduction to Particle Accelerators* course [S3-5]. It should be stressed in this present (hopefully!) late pandemic framework with several years of online courses in intensive school sessions, that there is strong consensus to return to in-person education. In **Sec. 3.4**, issues encountered with intensive online AS&E instruction are reviewed and strategies for framework improvements are given. These strategies are based, in part, on pandemic advances and lessons learned. The present vision here for online tutorials and MOOCs is not inconsistent with a return to in-person emphasis – there will be a persistent long-range need for augmented training and outreach that can fit a variety of contexts.

*3.1.3 PhD-level scientists and engineers:* This group is the backbone developing future concepts in projects to ensure the long-term health of our field. The process of educating the PhD-level accelerator workforce centers around universities and their partnerships with national laboratories and industrial programs. While the fundamental physics and engineering basics are generally provided in the university curricular, their specific applications to accelerator science and technology depends on links between faculty advisors and laboratory and industrial researchers. This aspect of student training is little different from that of PhD-seeking students in other sub-fields of physics and engineering. At the national laboratories the students and junior scientists are typically mentored by AS&E professionals who nurture their development and training within projects and/or by directed study.

Formal instruction in specialized AS&E topics is provided by intensive courses offered by the US Particle Accelerator School (USPAS) in the USA. Additionally, there are several regionally-focused, intensive accelerator schools around the world. These are presently most developed in Europe and include the CERN Accelerator School (CAS), the Joint Universities Accelerator School (JUAS), and the Nordic Particle Accelerator Project (NPAP) with linked MOOCs. The International Accelerator School (formerly JUAS) has also started with plans to convene a topical session every two years in a rotation of regions. Several accelerator conferences are offering mini-courses (a few hours) under the sponsorship of physics and engineering professional societies. The intensive-format course offerings of these schools are an essential supplement to what can be offered in university programs as universities do not have sufficient students to convene classes for only a few students. The status of the USPAS and community-identified opportunities for improvements are detailed in **Sec. 3.4**.

During the past four years, this model of university education plus intensive accelerator school courses has been augmented by four DOE-funded traineeships. These traineeships are run by universities, or collaborations of universities, and are each intended to launch ~4-6 university graduate students per year into areas of need within the accelerator workforce. Financial support for an individual PhD student within a traineeship is capped to 2 years, with this funding providing a start before the student is placed in a lab where they must be supported to graduation. Although these programs are helping bring in more students, constraints imposed open challenges. The status of the traineeships and community-identified opportunities for improvements are detailed in **Sec. 3.3**.

An instructional approach that could be considered for the USA is to adapt an approach taken in the United Kingdom by the John Adams Institute. In that case, a dozen or more entering graduate students from multiple universities are given a single design project to develop over a two-to-three-month period. At the end of the design period, the students present their work to faculty and external reviewers. In the US, such a program could be economically implemented by videoconference – which the community is now much more comfortable with.

*3.1.4 Core curricula:* The plethora of universities, traineeships, intensive schools and the wealth of approaches contained within them are a strength of the AS&E field. This diversity ensures that a broad range of technical approaches are considered and develops a consistent spectrum of talent. However, with this breath, there is considerable uncertainty what core skills should be consistent with levels of degree and specialization. The community might consider providing guidelines of core competencies for the universities and DOE Traineeships. Several traineeships are already providing certificates attesting to core AS&E competencies. Generalization could be considered.

**3.2 Graduate Education at Universities**

University-based research provides an important training ground for accelerator scientists and engineers. While the number of universities with faculty in accelerator physics and engineering is small, they provide excellent training, and the programs at Cornell, Indiana U, IIT, MIT, MSU, NIU, Old Dominion U, Stanford U, Stony Brook U, UC Davis, UCLA, U Maryland and other institutions have produced many of the leaders in the field. Importantly, these programs increase the visibility of AS&E among undergraduates, and summer research with these university groups, for example, through the NSF Research Experience for Undergraduates (REU) program, is an important pipeline into the field.

Facilities access is essential to university-based graduate student education. Many universities have local facilities, such as PEGASUS at UCLA; CESR, CBETA and MEDUSA at Cornell; and FRIB at MSU, where students have the opportunity for hands-on projects linked to facility innovation and upgrades. Other universities make use of test facilities at the national labs, such as AWA at Argonne, ATF at BNL, and IOTA at FNAL. Some students take advantage of the Office of Science Graduate Student Research program to access lab facilities and work with lab scientists. Access to facilities, particularly those on campus, drives cutting-edge research and helps university faculty – and AS&E as a field – draw in student talent.

The Center for Bright Beams (CBB), an NSF-funded Science and Technology Center led by Cornell, supports graduate student research at Cornell, U Chicago, Arizona State, Brigham Young, U Florida, Northern Illinois University, UCLA, and U New Mexico. Three national labs, SLAC, FNAL and LBNL, also participate. Since its inception in 2016, CBB has trained 20 PhD students and 11 postdocs who have gone on to careers in academia, in industry, at national labs, and at least 40 more will graduate by the time of CBB's sunset in 2026. CBB research is interdisciplinary, with approximately one half of its trainees specializing specifically in accelerator science.

It is critical to sustain university-based research, both for its graduate education and as a pipeline into the field. A sufficient number of funding calls and support are needed to provide continuity of research support for university faculty with adequate duration for graduate student projects. Importantly, sustained research support incentivizes universities to dedicate faculty lines to AS&E. The DOE GARD and Stewardship programs provide the majority of university support, with additional contributions from NP and BES. NSF provides support through CBB and its program in Plasma Physics. Other activities are directed to individual NSF programs benefiting from advances in accelerators, but the broad applicability discourages their support. DOE faces a similar issue, but has recognized the lead role of OHEP. Unfortunately, the NSF funding call in accelerator science was put on hold in fiscal year 2018, which sent a counter-productive message to the universities supporting the AS&E community.

In summary, we recommend that the community take the following step:
- Sustain the pipeline into AS&E by maintaining its visibility on university campuses. This requires sustained funding to faculty researchers, including support for on-campus accelerator facilities.

### 3.3 Department of Energy Traineeships

University education has recently been augmented by four DOE-funded traineeships supporting AS&E graduate students at universities. These programs are supported by 5 year grants and are designed to start ~4-6 university graduate students per year in the field with two years of capped funding eligibility. The goal of the funding is to provide a start for individual students before they transition to national laboratories and other programs to complete MS or PhD thesis research. The first traineeship was the Accelerator Science & Engineering Traineeship (ASET) funded in FY 2017 at Michigan State University (MSU) to convey (primarily PhD) physics and engineering students to various national labs. Two other traineeships were funded in FY 2020: the Courant Traineeship run by Stony Brook University and Cornell University linking students with Brookhaven National Lab; and the Chicagoland Accelerator Science Traineeship (CAST) run by Illinois Institute of Technology and Northern Illinois University to place MS and select PhD students at Fermi and Argonne National Labs, or regional industry. A fourth traineeship, the Virginia Innovative Traineeships in Accelerators (VITA) was funded in FY 2022 and is run by Old Dominion, Norfolk State, and Hampton Universities to place students in Jefferson National Lab. The VITA traineeship will emphasize underrepresented minority recruitment into the field and will also focus on MS education with select students elevated to PhD studies. A brief summary of the traineeships can be found at [2].

The support of the four DOE-funded traineeship programs (ASET, Courant, CAST, and VITA), are essential to building the future competency of the US AS&E field. Each traineeship works closely with affiliated universities to establish MS and/or PhD programs and also to attract qualified students from a background of physics, mechanical- and electrical-engineering. The individual traineeships have unique organizational challenges specific to the universities affiliated with each traineeship. However, the four programs share common challenges including: recruiting domestic students with the potential to be strong performers, placing students in laboratory research projects, and the continuity of program funding. Each of these are discussed below.

**Recruiting high-potential domestic students** is not only crucial for the individual success of the traineeship programs, but also for the future competency of the accelerator workforce in the US. Attracting students with the potential to develop into strong research scientists and engineers will help meet future demands of state-of-art R&D and enhance international competitiveness. University-based research in AS&E attracts students to the field and provides opportunities for undergraduates to participate in summer research, which often results in their choosing to go to graduate school in AS&E. The number of university programs is small, however, so many undergraduate students are not exposed to AS&E. Consequently, the traineeships need to attract domestic students from other appealing directions in physics and engineering. Other fields such as high-energy or condensed matter physics have a long-established university presence to act as a conduit of student talent. High profile topics like artificial intelligence, machine learning, and quantum computing have relevant aspects within the field, but students do not typically connect AS&E topics to them. Recruiting challenges in engineering can be equally or more severe. For example, undergraduate university courses in RF systems rarely emphasize aspects relating to core accelerator technology. Industry also has many high-paying jobs to lure away talent. A coherent community effort in AS&E education and outreach for undergraduate students is needed to increase the visibility of accelerator science and technology as an attractive career option. Favorable aspects can be emphasized: that ample jobs exist in the field and that interesting and achievable career paths accommodate a broad variety of applied, theory, and computational directions. Recruitment can be further enhanced by accelerator faculty/researchers participating in undergraduate summer research programs such as REU, SULI, and CCI, and actively advertising among newly-admitted graduate students who are seeking research opportunities. Field-wide recruiting materials could be updated and maintained (see **Sec. 4**) taking advantage of short videos and social media. Sustaining, or better increasing, university-based research is important for visibility. An additional junior-level undergraduate survey course affiliated with the USPAS (see **Sec. 3.4**) could also be developed to reach out to uncommitted undergraduates to make students aware of traineeship opportunities for graduate studies. Such a course should be a national program and might also generate outreach and tutorial materials (such as recorded lectures) for use in recruiting. Data on job opportunities within the field and long-term career paths of former traineeship students should also be collected to help recruit students. Impact of this might be particularly strong in physics since other historically attractive fields such as high-energy and nuclear physics have comparatively more difficult job markets. The traineeships should track former students (MSU and ODU are presently doing so) and we advocate a systematic community effort for gathering data on field job opportunities (see **Secs. 3.4** and **Sec. 4**).

**Placing students in laboratory research projects.** For an individual student, the traineeship awards currently provide capped funding for up to two years and $110k. If the student is supported as teaching assistants their first year while focused on core course requirements, the funding can be used as a bridge to support the student starting in the field before they are placed in laboratory or industry projects to continue research support till graduation. Usually students are sent to several USPAS intensive-school sessions for specialized training to help them further prepare for their research. The traineeships rely on relevant courses being available and the support of USPAS via scholarships (which do not fund travel) to support the students. This model, along with a limited summer traineeship period in a lab, can be sufficient support for MS students. For PhD students and MS students that decide to continue to a PhD, a traineeship can only cover a small fraction of overall support for the typical 4-6 year durations needed to complete a PhD degree. The remaining funding is expected from the local university research program, if such a program is in place, or at hosting groups in national laboratories or other traineeship partners that the student is placed in after initial university training. The availability and continuity of potential laboratory programs to host a specific student are often not well synchronized with the student's training and graduation timeline. Organizational issues to set up support can be substantial and labs may only want to commit to supporting the strongest students. On the other hand, the traineeship student may only be exposed to limited research opportunities, usually limited by the personal connections of the university faculties. Additionally, some students change plans and may withdraw in the middle of the program. It would help the traineeships if the DOE would make clear expectations of consistent student support within national lab programs, consider constructing dedicated funding mechanisms for pairing PhD students within the traineeships and the potential research groups, and recognize that it is not realistic to assume that all students will complete the programs after enrolling and may draw partial support.

**Continuity of program funding** is also highly important. The traineeship funding awards are for 5 years, much shorter than the typical time for establishing a reputable PhD program, which usually demands several Ph.D cycles (average 6.3 years in 2016-2017 [3]). The uncertainty of traineeship funds after 5 years renders the programs awkward for universities and many students recruited into the programs. Capped student support amounts of $110k are also small relative to the estimated cost to produce a PhD, which ranges from $300k - $400k or more, depending on the university cost, duration to degree, and topic. Several universities barred accelerator groups from participating in previous

traineeship proposals due to high unfunded liability per student resulting from the capped student expenditures and potential liabilities extending beyond the duration of traineeship funding awards. It would be prudent for the DOE to consider committing to a limited timeline of support for students till graduation for those admitted during the scope of traineeship funding awards. This would avoid unfunded liabilities and be fairer to students recruited to work in AS&E. The DOE should also clarify performance levels where the traineeships can have reasonable expectations of renewal so the programs can function long-term for maximum positive impact.

**3.4 US Particle Accelerator School**

Training for graduate students and early career professionals is often augmented by intensive school classes provided by the US Particle Accelerator School (USPAS). The USPAS was created in 1981 and has been operating under a university-style format since 1987 with pandemic adjustment to move online from Winter 2021 - Summer 2022. The USPAS focuses on educating graduate students and early career scientists and engineers in graduate-level courses. The courses are delivered in an abbreviated intensive format in two-week duration in-person sessions. Although the format is abbreviated, instructor contact hours substantially exceed the ~45 hours required for 3 semester hours of university credit conveyed. The USPAS typically (pre-pandemic) serves 300 to 350 students/year with 20 to 26 specialized courses per year delivered in two sessions (Winter and Summer). Details on the USPAS can be found at [1]. The curriculum of the USPAS covers the full field of accelerator physics and engineering, and evolves under the guidance of community-derived governance committees.

Pre-pandemic, from 2019 through 2020, the USPAS convened exceptionally large sessions due to high demand for accelerator training resulting from: healthy national laboratory accelerator programs with a plethora of new projects and facility upgrades, an aging workforce that has been retiring which results in new hires, and increasing opportunities for accelerator specialists in industry and medical applications. Post-pandemic, high demand for accelerator specialty training is likely to resume. Pandemic induced hesitancy toward online courses may drive higher rebound demand when in-person sessions resume. The pandemic forced a pause of in-person sessions starting from Summer 2020 through Winter 2022. The shift to online sessions has increased the effort required to convene sessions with many course reconfigurations and postponements. Cloud computing resources and other software became essential, and the USPAS office created extensive preparation sessions to assist instructors unfamiliar with online teaching to deliver their courses. This shift to more emphasis on software has increased workload for the school office with two full time workers and a ¾ time director.

The school must return to in-person sessions as soon as it is safe and practical to do so. Although reviews from online sessions have been strong (close to 100% of students rate their courses as being somewhat or very important to their future, teaching teams received high reviews, etc), students and instructors struggle to teach and learn within the online format. The lower third of classes has notably poorer online performance relative to in-person classes and reports high stress. Approximately three quarters of teaching teams have chosen to not teach online due to: schedule incompatibilities, the longer duration necessary, incompatibilities of the format for laboratory courses, and skepticism that online classes are worthwhile due the enhanced difficulty in maintaining the connections and networking benefits that in-person sessions convey. Since the USPAS is organized primarily around donated teaching effort from DOE labs, the USPAS cannot require instructors to teach. Numerous issues with teaming to deliver classes with shifting pandemic constraints must be addressed to allow specific classes to convene. Much needed applied classes with laboratories have been postponed. Our most basic class, *Fundamentals*, required prodigious effort to reconfigure essential measurement laboratories for online. Teaching teams that have taught a rotation online tend to view the enhanced load are unlikely to be open to multiple online iterations. Consequently, it will become increasingly difficult for the USPAS to convene viable online sessions if the pandemic extends into or beyond 2023.

Efforts to deliver online USPAS courses have been formulated to bring long-term benefits to the AS&E community in the eventual resumption of in-person sessions. Cloud software tools that are essential for online classes will remain central in post-pandemic, in-person sessions. These include: cloud PCs with preloaded software and browser accessible services such as the RadiaSoft *Sirepo* GUI and linked *Jupyter* server that support a wide variety of accelerator codes; use of *Zoom* to allow bringing in key lectures with more flexibility and potentially recording/archiving in-person lecture streams for augmented class materials to increase impact; use of cloud file sharing tools such as *SharePoint, Google Drive,* and *Dropbox*, enhanced website services, classroom management software such as *Google Classroom*, *Canvas*, etc to help organize teaching efforts; networking software such as *Slack* and *Discord* to archive class technical discussions that can extend beyond classes; and interactive electronic whiteboards to save and enhance aspects of lectures, etc. Cloud-installed beam simulation codes on *Sirepo/Jupyter* are provided by a support contract with

RadiaSoft LLC. These RadiaSoft software tools require no setup (browser accessible) and remain freely available for use by students after their sessions – thereby bringing additional value to student research and lab projects. Additional cloud computers/workstations (presently via *Amazon Web Services* windows or linux workstations) are accessed by a simple to install app (available on all common platforms including smartphones) on student computers and accessed over the network. These cloud computers provide considerable flexibility and allow instructors to better prepare in advance of sessions. Feedback from students and instructors on cloud computing has been strong through the two online sessions in 2021. However, reliance on more software and cloud-based resources will necessitate a higher degree of local network evaluation/verification for future in-person sessions than was the case for pre-pandemic sessions that relied on rented physical computers set up a day or two in advance of the sessions.

These software and network aspects result in increasingly technical/IT issues in future in-person sessions. Additionally, shared codes and increasingly open-source software and libraries of simulation tools used by the field opens opportunities for our classes. The school should be active in encouraging and facilitating transitions of legacy codes to open source with efforts to update interfaces to create useful tools for both our classes and the community. It is anticipated that present USPAS office staff of two FTEs will be highly loaded in the expected return to two larger sessions/year due to the load to setup larger session venues, academic credit, housing and travel needs for larger student counts and teaching teams, etc. It is prudent to augment office staff with an individual with IT skills to deal more effectively with this extended load. Added staff would enable additional tasks to benefit the community, and allow for long-range and contingency planning.

The USPAS has recently enhanced diversity, equity, and inclusion efforts in response to growing emphasis on these issues. Data on women and underrepresented minority students in USPAS is presented in **Sec. 5.** The data shows the USPAS doing relatively well compared to the broader AS&E field. But there is room for further improvement. Efforts have continued to promote the role of early-career professionals, women and underrepresented minorities on teaching teams. This provides role models and is more welcoming to underrepresented groups. While percentages of women students (21%) and instructors (15%) in recent 2021 sessions are likely somewhat higher than the composite field and has risen steadily from 1987, overall progress has been slow. Enhanced recruiting efforts will likely be necessary to further accelerate the fraction of women in AS&E. Efforts to recruit underrepresented minorities have been recently enhanced. A new *Sekazi Mtingwa Scholarship,* named in honor of the 1st African American Wilson Prize winner Sekazi Mtingwa, was instituted in Winter 2022 to provide full (including travel) support to qualified underrepresented minority students in USPAS sessions, thereby reducing potential financial obstacles to attend. This effort has received broad attention. Prof. Mtingwa has agreed to recruit students for this scholarship. Additionally, the USPAS will target a historically black university, Florida A&M University (FAMU) to act as academic host for a session post pandemic. FAMU has a sister-field program in plasma simulation and diagnostics that should result in a natural pairing for long-term mutual benefit.

To significantly increase participation of underrepresented minority students, particularly those of Black/African ancestry, recruiting efforts must be expanded. Long-term statistics of minority involvement throughout AS&E are also sorely needed to guide the USPAS and the accelerator community efforts and to measure progress. To be more broadly welcoming, the USPAS has: codified a diversity and inclusion policy, a code of conduct outlining expectations of respect and professional conduct, and carefully updated collection of statistics related to gender and ethnic identity. The USPAS also regularly promotes highlights of people from historically marginalized groups on the school website and on social media (USPAS *Facebook/Meta*). Targeted advertising to Minority Serving Institutions has been expanded and increased.

The USPAS has been responsive to the DOE Traineeships. All four traineeships are represented in USPAS governance functions including the *Curriculum Committee* guiding our choice of courses. This emphasis should be maintained. The school has begun tracking participation of traineeship students in USPAS classes. Additionally, the USPAS should request written input yearly from the traineeships on specific course needs, so the school curriculum can best respond to specialized training needs. Enhanced AS&E recruiting of undergraduate students from a more basic survey course should also benefit the traineeships. Additional aspects of needs for the DOE Traineeships are discussed in **Sec. 3.2**.

Enhancing cooperation with the various international schools is logically appealing in light of the large fraction of the domestic effort originating from international talent originating from other countries. Countering this sentiment, there is increasing government scrutiny of international talent recruitment efforts in domestic universities, labs, and industry. It is usually impractical to support domestic students traveling abroad to attend international schools. The USPAS continues to support a relatively low percentage (~10%, primarily European) of international students to attend our sessions. Many within this group of students are linked to collaborations with US projects, are very strong students that enhance our classes, and have a high rate of seeking employment at US institutions after they receive their degrees. The USPAS also continues to coordinate with the biennial Joint International Accelerator School. We recommend that the USPAS maintain these efforts. Additionally, the USPAS should explore opportunities for international cooperation on tutorials and archives of class materials to ease production of quality classes.

The following enhanced roles for the USPAS would strongly benefit the AS&E community. The present office staff is heavily committed in ongoing school operations. So these extra roles would require augmented effort to implement, and in some cases, additional funding for software and extra sessions. Depending on the level of augmented effort and priorities, some subset could be prioritized.

- **Tutorials:** Coordinate with conferences such as NAPAC, IPAC, CARRI, etc to regularly generate tutorial materials to advertise USPAS courses, attract students, and for the community to update skills.
    - Pre-pandemic experiment with this model linked to NAPAC 2019 in Lansing was successful and verified the merit of the approach.
    - Elementary tutorials and effective distribution should aid recruiting.
    - Opportunities should exist to coordinate production of tutorial materials with international accelerator schools.
- **Community Statistics:** Track community hires and career paths of former students in labs, academia, and industry to support USPAS course decisions and justify the DOE Traineeships. Track gender and other demographic diversity in labs, academia and industry to help guide efforts to improve equity within AS&E.
    - Pre-pandemic experiment collecting 2020 hire intent from USPAS collaboration labs demonstrated feasibility and usefulness of the data.
    - Could exploit software such as *Qualtrics* to assist packaging results for optimal community use.
- **Improved Dissemination of Course Materials and Augmentations:** Via enhanced cloud storage interfaces and instructor templates/assistance to improve distribution and use of course materials.
    - Pandemic shows video records of lecture streams enhance course impact. Editorial assistance is needed to improve formulation packaging of such materials.
- **Instructor Resources Library to Ease Course Formulation:** Courses involve much work to formulate that is often part repeated. It would improve classes for USPAS to maintain resource libraries for problems, common topics, code tools, etc to assist instructors making their courses.
    - *Fundamentals* laboratories are now being effectively shared by four teaching teams to improve quality and reduce workload.
    - *LaTeX* problem library initiated and successfully used in a few classes to reduce work for creating vetted problems.
    - Use of open source simulation tools and libraries in our courses can be facilitated to the broad benefit of the community.
    - Common graphics, programs, etc. could be maintained.
    - Opportunities exist to coordinate this effort with international accelerator schools such as CERN and JUAS.
- **Enhanced Recruiting:** USPAS sessions focus on graduate and early career professional students already specializing in AS&E and are not designed as a light introduction to draw in undergraduate student recruits. It is essential to consistently attract students with high potential for the future health of AS&E. Scheduling a yearly junior-level undergraduate course could draw in additional student talent, make them aware of opportunities in the field, and help launch them on favorable pathways. This course should be targeted for both physics and engineering students, more accessible than USPAS *Fundamentals* (typically taught at senior-level undergraduate or beginning graduate student level), and serve as an overview highlighting the breadth of science and technology opportunities in the field.
    - Present session sizes (7-8 concurrent classrooms) are fully committed and cannot be expanded without much higher cost. Time phasing and coordination with other school activities would also be important. Options would need to be carefully evaluated.
    - Help recruit domestic students into the DOE Traineeships (see **Sec. 3.2**).
    - Enhance recruitment of women and underrepresented minorities (example: advertise at and convene sessions at HBCUs).
- **Enhanced Social Media Presence:** The school has benefited from outreach on *Facebook/Meta*. Additional approaches could be tried to enhance recruiting, maintain awareness of educational resources and career opportunities, and build community.

- **Pandemic Catch Up:** The pandemic resulted in a large backlog (27+) of deferred specialty courses that are now unscheduled. It would benefit the community to schedule a 3rd smaller session for a transient period to reduce the backlog of specialty courses.
    - 2-3 year teaching cycles may be dilated to 3-5 years, resulting in potential loss of valuable classes.
    - An additional smaller session, 1x per year with 2-3 classrooms, would help restore teaching cycles while being easier/cheaper to find venues and organize.
    - Session topics and locations could be chosen to advance specific themes.

Additional USPAS office effort to realize part or all of the extended roles above requires ½ (min practical) to 1 FTE and would cost (Fermilab overhead included) ~$90k to $180k per year. The individual would need IT/software skills to assist with increasingly technical aspects, have good people skills, be open to at least twice-yearly extended travel to the sessions, and should be early-career or view the position as a long-term opportunity. Hiring a dedicated FTE would simplify searching for the appropriate person relative to trying available part-time effort at Fermilab. This path also addresses long-term planning needs. The present USPAS manager started tenure in 1984, has worked with all directors to develop strategies for the school, and is expert in the multitude of details involved in operating the school. The USPAS has strongly benefited from this exceptional continuity, high-level performance, and dedication. It would be wise to bring in additional help while the office is stable and high functioning to allow extending school missions to the benefit of the community while providing appropriate contingency and long-term planning.

## 4. OUTREACH

AS&E outreach activities must be extended to help bring in talent to secure US leadership in future R&D directions. The DOE traineeships (see **Sec. 3.2**) are also struggling to recruit domestic students, and would benefit from an enhanced community-wide effort to attract talented individuals into relevant undergraduate and graduate educational programs. Recruitment of such talented individuals is needed to secure the technical future of AS&E. Approaches to outreach should evolve based on the ways in which high school and undergraduate students are influenced, with emphasis placed on raising awareness of exciting technical and career opportunities in AS&E. Present outreach efforts include the Education, Outreach and Diversity (EOD) subcommittee of APS DPB (Division of Physics of Beams). This committee is working to update a division community outreach and recruiting brochure that was last updated in 2013. The APS DPB executive committee provides travel scholarships for students to attend USPAS sessions and the APS April Meeting. In this section, we review these efforts, propose refinements, and suggest additional approaches to improve outreach and recruiting.

The 2013 APS DPB brochure is being updated to reflect changing priorities within the field, and generational values including contributions to climate change research, COVID-19 pandemic research, internationality, and diversity. This update is targeted for distribution in late 2022. It is intended to raise awareness of AS&E opportunities, and potentially serve as a recruitment tool for students and faculty unaware of the field and its contributions. The brochure update will be designed for accessibility by advanced high school and undergraduate students, for broad distribution including university physics departments. Distribution was previously limited to printed versions and APB DPB website downloads [S4-1]. We plan to diversify distribution avenues via social media, community websites, YouTube videos, Wikipedia links, and other outreach avenues.

APS DPB travel awards (~24/year anticipated post-pandemic) supplement USPAS financial aid (which generally does not include travel support) to fill a need while encouraging students to join DPB. This membership is critically needed to boost low student participation, and to the field maintain APS division status. As of January 2022, DPB has 2.23% (1056 members) of APS membership, which is only slightly above the 2.2% required for division status [S4-2]. Loss of division threshold for DPB would reduce the status of the DPB to a Topical Group, lower (or eliminate) APS fellowships from DPB, reduce the number of APS sessions allotted to the our field, and thereby hinder future talent recruitment. The present fraction of student members in APS DPB is 13.6%, which is much less than other APS Divisions which range from 25.8% to 54.1% [S4-2]. Even small increases in student membership should significantly help this situation.

Undergraduates who are at universities without programs in accelerator science seldom envision themselves working in AS&E. They must be offered an intriguing and achievable view about the excitement and high impact of AS&E commensurate with their work-life ambitions. Approaches to draw in talent will be more productive if they focus on junior-level undergraduates who have not yet been locked into specific fields of study. This is also an effective age to reach women and underrepresented groups to try and address longstanding equity and inclusion issues (see also **Sec. 5**). Possible approaches include:

- Initiate a national junior undergraduate-level recruiting course to alert students to the breadth of the field and range of technical opportunities in physics and engineering.
    - Potentially run by the USPAS (see **Sec. 3.4**).
    - Must be introductory to be accessible to students unfamiliar with the field and reach junior-level students in both engineering and physics.
    - Should also recruit students for APS DPB enrollment.
- Produce more high-quality basic field overview videos, and distribute in a manner where students yet to make career choices might be influenced.
    - Could be coordinated with the recruiting course described above.
    - There are a number of ongoing efforts [S4-3 through 8] which need to be coordinated.
    - A community-wide effort to improve relevant Wikipedia articles would also be very beneficial.
- Accumulate field job and hiring statistics to help students understand there are interesting and achievable career paths.
    - Could be taken up by APS DPB or the USPAS in extended roles (see **Sec. 3.4**), or some combination thereof.
    - Distribute data to DOE Traineeships (see **Sec. 3.4**) to help them draw in students.
- Consistently support university-based research in AS&E to draw in more student talent.
- Members of our community should be incentivized to participate in undergraduate level outreach. For example, meeting with or presenting to local chapters of the society of physics students (SPS). Our institutions should be encouraged to officially support such outreach efforts.

## 5. Diversity, Equity, and Inclusion

For long term health of the field, the accelerator community must improve its performance with respect to diversity, equity, and inclusion. Limited data from recent USPAS sessions (see **Fig. 5.1**) suggests that the fraction of women attendees grew from ~8% in 1987 to ~20% in 2016, then plateaued within a 5% band. Representation of women in DOE national labs as a whole is roughly 31% across all departments and divisions [S5-1]; when narrowing the focus to research and technical staff and management positions, women make up less than 20% and 19% respectively. When considering AS&E focused divisions or departments, the percentage of women at DOE national labs is somewhat lower than USPAS student numbers with a heterogeneous distribution among the labs and subspecialties [S5-2 through 8].

Several well understood barriers inhibit gender diversity in STEM fields as a whole. These barriers include: lack of access to affordable childcare and eldercare facilities, respective maternity leave policies, toxic work enviornments, de facto discriminatory recruitment and hiring practices [S5-9], and gender-based microaggressions and harassment [S5-10 through 13]. Accelerator physics and engineering is not immune to these issues, which negatively impact hiring and retention of underrepresented groups. For example, DOE laboratory programs are often not sufficiently accommodating for family needs such as childcare. Although this issue negatively impacts both men and women at work, the impact is statistically more significant for women [S5-14, S5-15]. Such issues slow progress in increasing the percentage of women in our field. Leading examples of AS&E diversity should be studied to document best practices. For example, more than half of current postdocs at the Cornell Center for Bright Beams are women.

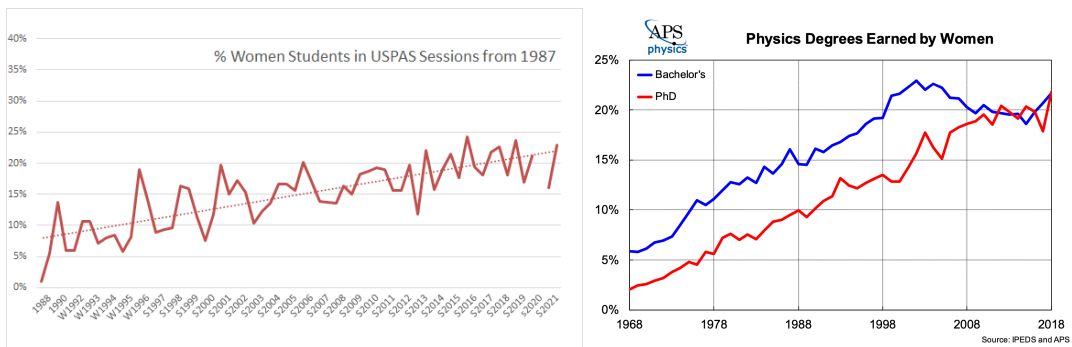

**Fig 5.1** Fraction of USPAS women students in sessions since 1987 and number of physics degrees awarded to women since 1968 [S5-16]. USPAS statistics tracks with the wider physics community.

Many barriers to access and inclusion also significantly impact the participation of people of all genders from historically underrepresented minority groups (URM) [S5-17]. This includes people who identify as Black or African American, Hispanic or Latino/a/x, and American Indian or Alaska Natives. Prior to 2021, USPAS was not collecting race or ethnicity data of students consistent with URM classification. For the three most recent USPAS sessions where data is available, self-reported URM fractions properly accounting for mixed heritage indicate useful URM fractions (with absolute Black/African American percentages): Winter 2021 16% (2 students), Summer 2021 14% (1 student), and Winter 2022 15% (5 students). The URM student percentages are dominated by students of Hispanic ancestry. However, counts of students with self-reported Black/African American or Native American ancestry are very low and were known to be chronically low prior to 2021. The relatively high Black/African American count for Winter 2022 may be due to targeted recruiting efforts in USPAS and the labs, a new URM scholarship started in Winter 2022 (see **Sec. 3.3**), and/or statistical fluctuations. That level is unlikely to be sustainable long-term without further enhanced recruiting efforts of URMs with Black/African American ancestry. Course instructors from URM groups are also problematically low, which also hinders recruitment due to lack of similarity mentoring. This trend continues at the national lab level with no lab reporting more than 13% minority representation lab wide [S5-1 through 8].

Missing from our data analysis is a statistical view of the intersection of marginalized identities in USPAS and the AS&E community more broadly. For example, while we know that more than half of aggregate science and engineering degrees at the bachelor's, master's and doctoral levels for URM students went to URM women [S5-17], we do not know if that same trend holds true for our field. We must collect data that allows us to understand the intersection of race, ethnicity, and gender as it pertains to participation and persistence in our fields and programs in order to develop appropriately targeted interventions.

The relatively slow progress in our field with the representation of women and URMs, particularly people of Black/African American and Native American ancestry, is sobering. An issue compounding efforts to target improvements is that data to measure the fraction of women and underrepresented minorities in the national labs and the broader AS&E community is overall sparse. Reliable and consistently gathered diversity data are essential to evaluate the successes and failures of present and future efforts toward a more diverse workforce. We advocate that a community wide effort be initiated to monitor demographic diversity statistics of AS&E workers in the national labs, universities, and private sector employers. This might be accomplished with needed long-term consistency by the enhanced roles proposed for the USPAS in **Sec. 3.4** and coordinated with ongoing lab monitoring efforts.

In addition to improved data collection efforts, several national labs (ANL, LBNL, LLNL, LANL, SLAC), have joined the APS Inclusion, Diversity, and Equity Alliance (IDEA) initiative. APS-IDEA was created with the goal "of empowering and supporting physics departments, laboratories, and other organizations to identify and enact strategies for improving equity, diversity, and inclusion (EDI)." [S5-18] Recent work taking place with the SLAC APS-IDEA team includes a comprehensive review of the internship programs and recommendations for improvement that range from recruitment strategies to application review. We recommend AS&E members at institutions with an APS-IDEA team learn from the research and resources that such groups can provide. If there are questions related to DEI efforts and best practices, members of an APS-IDEA team can prove a great resource.

Some national labs also have fellowships supporting the hiring and retention of minorities and women. The Al Ashley fellowship [S5-19] at SLAC was started in 2011, and doubled in size in 2021. While the program is modest in size (a maximum of 4 positions per year), it is an example, along with the Mtingwa scholarship, on how concerted effort can have an impact. To date, the Al Ashely program has a retention rate of 79%. ANL's Walter Massey Fellowship [S5-20] was founded in honor of the lab's first Black director, and 17 graduate students from URM groups were hired through the Graduate Education for Minorities (GEM) [S5-21] program in 2021 . Fermilab has recently initiated a similar program: Accelerator Science Program to Increase Representation in Engineering (ASPIRE) [S5-22]. With the first round of hires expected to start in 2022. Note this is not an exhaustive list, but rather some examples that result in some AS&E hires.

In addition to lab and institution wide efforts, there are several steps individuals can take to contribute to inclusiveness at their institution. For those at institutions with an outreach office, we recommend volunteering for a middle or high school event once a year. Often email notices in search of volunteers are sent well in advance of the events, and most events are broken into manageable shifts. Consider the impact you may have on a child's life by introducing them to AS&E. Another actionable, but not very time-consuming action is to evaluate the use of gendered language in your writing. If you are writing a course description or job posting, consider removing the use of he/she for a more neutral phrase. If you have more time to commit to DEI efforts, consider diverse hiring when selecting interns and staff. Consider hiring a community college student and/or consider other factors aside from GPA, test scores, and institution reputation. While test scores are often used when reviewing graduate school applications, studies show they are not always the best indicators of future success [S5-23, S5-24]. We recommend considering a variety of factors when hiring at any level, with less emphasis on traditional metrics, such as network hiring. For example, postdoc hiring tends to be based on the network of the hiring manager and warm outreach approaches rather than postings in channels that may encourage a wider range of applicants.

In summary, we recommend that the community pursue the following efforts to make progress in diversity, equity, and inclusion issues within AS&E:
- Gather integrated statistics on gender and ethnicity for AS&E students and workers in the labs, universities, and industry to monitor progress and guide long-term efforts.
    - Extended roles for the USPAS discussed in **Sec. 3.4** could fit this need well.
- The AS&E field should run a yearly undergraduate level recruiting program structured to draw in women and underrepresented minorities.
    - This could be coordinated with the USPAS (see **Sec. 3.4**) and the DOE Traineeships (see Sec. **3.3**)
- Strengthen connections to professional societies serving issues on diversity.
    - Example: explore collaborations addressing URM issues with the National Society of Black Physicists (NSBP) and the National Society of Hispanic Physicists (NSHP).
- AS&E should target more scholarship and fellowship support to draw in women and URMs.
    - The Fermilab ASPIRE program started in 2021 and the VITA DOE Traineeship (see **Sec. 3.3**) provide models.
    - The SLAC Al Ashley fellowship started in 2011 and has hired 19 fellows with a retention rate of 79%. Funding was recently increased for the program to support 4 fellows vs. 2 in the coming year (2022).
- APS-IDEA teams at several national labs, universities, and institutions are making recommendations to the physics community on DEI efforts. This work should be leveraged by the AS&E field.
- Encourage labs to reward employees for: volunteering for outreach and inclusion efforts, hirings addressing diversity, and efforts to enhance community welcomeness to underrepresented groups.